# Enabling Highly Efficient Solar Thermal Generation with 800°C-Stable Transparent Refractory Aerogels

Zachary J. Berquist,[1#]
Andrew J. Gayle,[2#]
Neil P. Dasgupta,[2,3]*
Andrej Lenert[1]*

[1] Department of Chemical Engineering, University of Michigan, Ann Arbor, MI 48109
[2] Department of Mechanical Engineering, University of Michigan, Ann Arbor, MI 48109
[3] Department of Materials Science and Engineering, University of Michigan, Ann Arbor, MI 48109

[#] These authors contributed equally
*E-mails: ndasgupt@umich.edu; alenert@umich.edu

## Abstract

Although spectrally selective materials play a key role in existing and emerging solar thermal technologies, temperature-related degradation currently limits their use to below 700ºC in vacuum, and even lower temperatures in air. Here we demonstrate a solar-transparent refractory aerogel that offers stable performance up to 800ºC in air, which is significantly greater than its state-of-the-art silica counterpart. We attribute this improved stability to the formation of a refractory aluminum silicate phase, which is synthesized using a conformal single-cycle of atomic layer deposition within the high-aspect-ratio pores of silica aerogels. The transparent refractory aerogel achieves a record-high receiver efficiency of 77% at 100 suns and an absorber temperature of 700ºC based on direct heat loss measurements at this temperature. Such performance and stability can enable the use of advanced supercritical $CO_2$ power cycles and lead to a ~10% (absolute) improvement in solar-to-electrical conversion efficiency. Transparent refractory aerogels may also find widespread applicability in solar thermal technologies by enabling the use of lower-cost optical focusing systems and eliminating the need for highly evacuated receivers.

**Keywords:** aerogel, concentrated solar thermal, spectral selectivity, atomic layer deposition

## Introduction

Developing receivers with high solar absorptance and low thermal emittance (*i.e.*, selectivity) that can endure temperatures above 700ºC could enable the deployment of more modular and less expensive concentrated solar thermal (CST) plants.[1–3] High temperatures are specifically needed to drive advanced supercritical $CO_2$ ($sCO_2$) power cycles, which are more efficient and compact than the steam Rankine cycles currently in use.[4] Such high temperatures also satisfy the demands of industrial processes that are otherwise difficult to decarbonize.[2,5–7] Although many approaches are being pursued to reach this high-temperature goal, line-focusing concentrators, such as parabolic troughs, offer distinct advantages, including lower capital costs and higher optical efficiencies compared to central receivers.[8,9] However, realizing efficient conversion of moderately concentrated sunlight (<100 suns) into high-temperature heat (>600ºC) remains challenging because of a lack of selective receivers that can endure these temperatures.

Current receivers achieve selectivity through thin optical coatings on durable, but otherwise not selective, materials such as metal alloys. These coatings were not designed and are unsuitable for emerging high-temperature applications. Temperature-related degradation, including oxidation, interdiffusion, and delamination currently limits their use to temperatures below 700°C in vacuum, and 600°C in air.[10,11] An alternative to selective coatings are transparent insulating materials (TIMs) which act to insulate a hot solar absorber (*e.g.*, a Pyromark-coated tube) from thermal radiative and conductive losses while allowing concentrated sunlight to reach the black absorber surface. Solar transmittance, thermal (infrared) opacity, and thermal conductivity are important properties in determining the performance of TIMs at elevated temperatures. Among TIMs, silica aerogels currently have the best solar transmittance and heat-insulating properties. The application of transparent silica aerogels in solar thermal systems at moderate/low temperatures has led to substantial gains in overall performance.[12,13] However, at high temperatures (>~600°C), conventional silica aerogels are not stable and do not block the infrared (IR) rays that account for most of the heat losses.[14,15] To address the latter limitation, our previous work has shown that introducing nanoparticles with plasmon resonances in the IR reduces heat losses at 700°C by a factor of two.[16] Nonetheless, silica aerogels undergo sintering and densification at elevated temperatures which degrades their solar transmittance and thermal conductivity,[17] thereby limiting their potential use in next-generation CST applications. Past efforts to stabilize or reinforce the mesoporous structure of aerogels, which is essential to their performance, have resulted in significant compromises in transmittance or thermal resistance.[18–20] Overcoming this challenge would unlock the full potential of moderate-concentration CST configurations and ultimately lead to large improvements in cost and reliability.

Here, we demonstrate a refractory aerogel that overcomes the thermal stability limitations of silica aerogels while preserving their key functional properties, namely high solar transmittance and high thermal resistance. We synthesize these aerogels using a recently developed process[21] for conformal, single-cycle atomic layer deposition (ALD) within the high-aspect-ratio pores (>60,000:1) of silica aerogels. The ALD process results in the formation of a refractory aluminum silicate phase which lowers the linear shrinkage rate at 800°C by an order of magnitude compared to the baseline silica aerogel. We thus refer to the modified aerogels as "refractory", in line with the ASTM definition, which specifies that refractory materials are suitable for applications that are exposed to environments above 538°C, and consistent with previous uses of the term in plasmonics.[22,23] Notably, owing to the large surface area of aerogels, our process introduces a large fraction of aluminum and significantly alters the overall chemical composition of the material, resulting in a Si/Al ratio ~ 3:1. Furthermore, based on direct heat loss measurements at 700°C with a transparent refractory aerogel on a SiC surface, we show that these aerogels enable the highest solar-to-thermal receiver efficiency at 700°C and 50-100 suns compared to existing solar-selective approaches that have been directly evaluated at ≥700°C. As a result of the improvements enabled by the ALD modification, this material has the potential to serve as an important component in next generation CST plants.

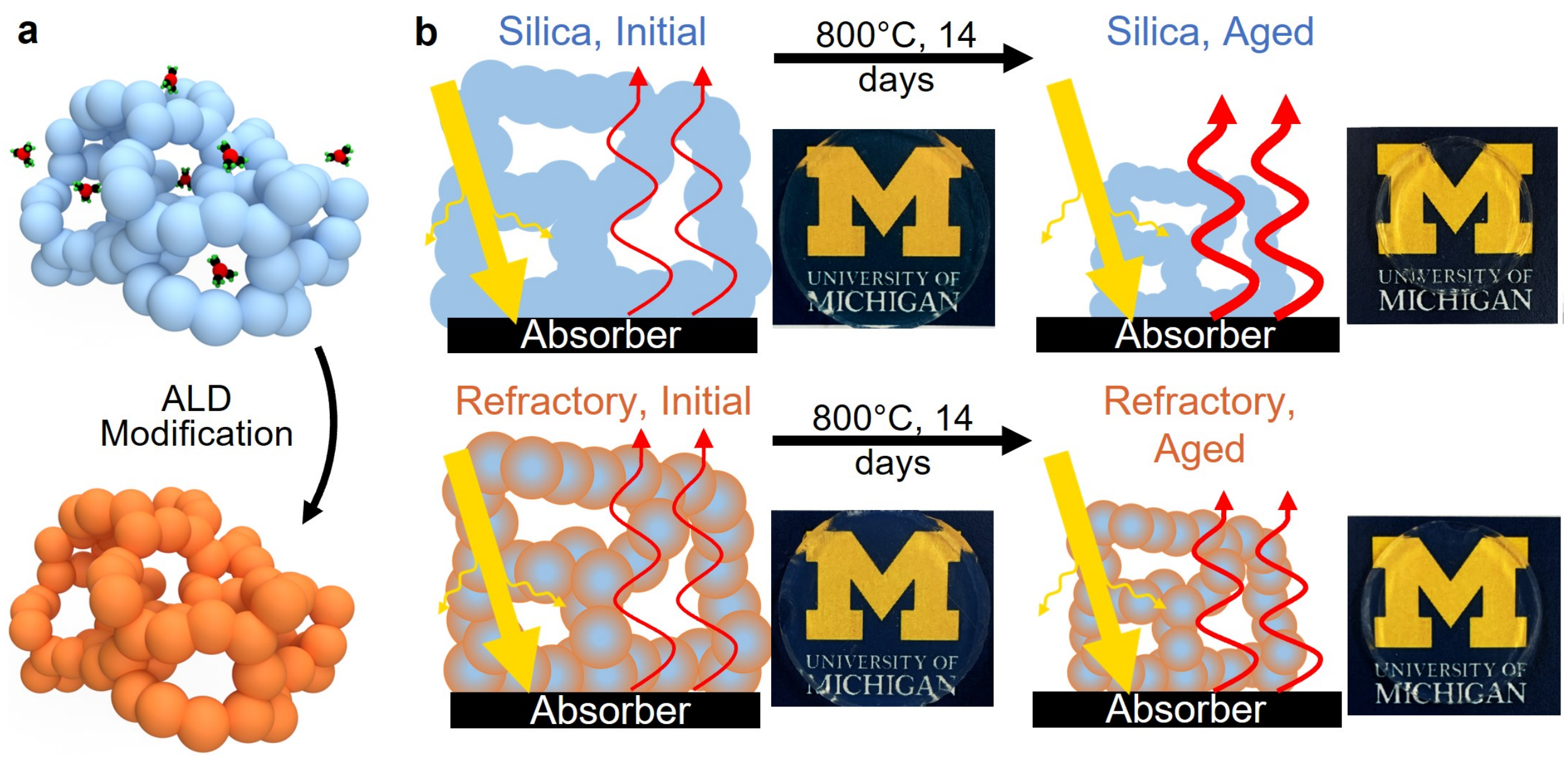

**Figure 1. Transparent refractory aerogels for solar thermal technologies. (a)** Schematic of silica aerogels following single-cycle ALD modification, resulting in the formation of a refractory aerogel. **(b)** Photographs and schematic illustrations of the aerogels before and after ALD modification and aggressive aging (14 days, 800ºC, in air), highlighting the solar transmittance and relative thermal stability of the refractory aerogels, which enables superior thermal insulation after aging. For reference, the height of the Michigan "M" is ~1.4 cm.

# Methods

## Aerogel synthesis

Aerogels were synthesized using our previously published procedure.[16] Synthesis was performed *via* a sol-gel polymerization of tetramethylorthosilicate (TMOS, 218472, Sigma-Aldrich) with deionized (DI) water and ammonia ($NH_3$, 341428, 2.0M in methanol, Sigma-Aldrich) as the catalyst. We first prepared two solutions. TMOS is diluted in methanol (MeOH, 322415, Sigma-Aldrich), and ammonia solution is added to DI water. The two solutions are then mixed in plastic syringes where gelation can proceed. Once the wet gels formed, they were aged for 1 week with daily ethanol washes to remove contaminants and water. The aerogels were then dried using $CO_2$ critical point drying (CPD, model 915B, Tousimis).

## Atomic layer deposition

ALD modification of aerogels was performed in a custom hot-walled, cross-flow ALD reactor.[24] Detailed methods can be found in our preceding publication.[21] Trimethylaluminum (TMA, 663301, Sigma-Aldrich) and DI water were used as precursors. Both precursors were kept at 37°C, and the deposition temperature was 150°C. Purging was performed using Ultra High Purity Ar (99.999%). A multidose-quasi-static-mode approach was used.[21] The exposure time was 400 s, followed by a purge time of 800 s. TMA was dosed 45 times, followed by 45 DI water doses. One cycle of ALD was performed in each deposition.

Before performing ALD, the aerogels were held at a vacuum pressure of ~600 mTorr under constant Ar purging for at least 1 hour at 150°C to ensure removal of excess water and other adsorbed species. This was followed by 5 DI doses to ensure hydroxyl termination of the aerogel surface.

## XPS measurements

To prepare samples for x-ray photoelectron spectroscopy (XPS), the aerogels were crushed using a mortar and pestle to create a powder. A mullite reference powder (2288122, Alpha Aesar) was also analyzed. A bulk ALD $Al_2O_3$ control was prepared on a planar Si substrate, using the same ALD instrument described in the Atomic layer deposition methods section. A constant-flow mode was used. The pulse times for TMA and DI water were 0.05 and 0.1 s, respectively, and the purge time was 30 s. 150 cycles were performed, resulting in a film thickness of ~21 nm.

Measurements were performed using a Kratos Axis Ultra XPS with a monochromated Al Kα x-ray source (10 mA, 12 kV). The spot size was 700 μm x 300 μm. An electron gun was used to maintain charge neutrality on the surface of each sample. Survey scans (pass energy: 160 eV) were used to quantify the atomic composition of the various samples. Core scans (pass energy: 20 eV) were used to investigate the binding environment of elements in each sample. The binding energies are calibrated to that of adventitious surface carbon (284.8 eV).

**Thermal aging**

The aerogels were annealed in a tube furnace (MTI Corporation) with a 2°C/min ramp rate in ambient conditions. The holding times for each temperature exclude the time to ramp up and down. The samples were aged under two different conditions. For the first, the aerogels were annealed at 700°C for 24h, followed by two 7 day anneals at 800°C. The aerogels were cycled down to room temperature after each anneal. For the second test, the samples were aged at 700°C with cycling down to room temperature after 1, 2, 4, 7, and 10 days. This process was repeated at 800°C for a total of 20 days.

**Density and porosity measurements**

The volume of the aerogel monoliths was calculated from multiple measurements of the diameter and thickness of the disks. The mass of the aerogels were measured using a standard analytical lab balance. To calculate the porosity, the amorphous densities of both alumina and silica were used to calculate the volume consumed by the aerogel backbone.

**Surface area measurements**

Surface area was measured using a Micromeritics ASAP 2020 Surface Area and Porosity analyzer. Prior to analysis, the aerogel samples were degassed at 350 ˚C (achieved by a ramp of 20˚C/min) for 8 hours. Brunauer–Emmett–Teller (BET) surface area analysis was conducted at 77 K using the amount of nitrogen adsorbed at various relative vapor pressures between 0.05 $< P/P_0 < 0.3$. Total pore volume measurements were taken at a relative vapor pressure, $P/P_0$, of 0.995. The molecular cross-sectional area of nitrogen for the analysis was assumed to be 0.1620 $nm^2$.

**Optical property measurements**

The solar direct-hemispherical transmittance of the aerogels was measured with a UV-Vis-NIR spectrophotometer (Shimazdu UV-3600 Plus) and an integrating sphere attachment (Shimazdu, ISR-603). The infrared transmission measurements were obtained with a Fourier transform infrared (FTIR) spectrophotometer (Fisher Scientific Nicolet iS50).

**Heat flux and thermal conductivity measurements**

The aerogels were placed on a custom hot stage (Instec) with a high-precision temperature controller (Instec MK2000) in vacuum. The heat flux was measured with a heat flux sensor (greenTEG gSKIN XP 26 9C) covered with a blackbody absorber (Acktar Metal Velvet). The heat flux sensor was adhered to a copper mesa block which was actively maintained near room temperature using compressed air.

The thermal conductivity of the aerogels was measured using the transient plane source (TPS) method (Hot Disk TPS 2500 S). Kapton sensors (Hot Disk Kapton 5465) were used with a heating power of 5 mW for a duration of 40 seconds. A correction factor associated with highly insulating materials was applied.[25]

**TGA measurements**

Aerogels were ground up with a mortar and pestle and then placed in a TA Instruments Q500 TGA under a flowrate of 50 mL/min of pure nitrogen and 50 mL/min of dry air. The samples were then heated at 5°C/min from room temperature to 150°C and held for 1h. The samples were then heated again at 5°C/min to 700°C where they were held for 1h. The samples were then allowed to cool to room temperature.

**DRIFTS measurements**

The aerogel samples were crushed up into a powder using a mortar and pestle. The aerogels were then dispersed into a potassium bromide (KBr) powder with the aerogel consisting of ~5% by mass. A Thermo Fisher Nicolet IS50 FTIR was used with Pike DiffusIR DRIFTS attachment. The DRIFTS accessory was purged with nitrogen. For the temperature-dependent measurements, the aerogel samples were held at each temperature for at least 30 minutes until the spectra remain unchanged from water desorption.

**Emission measurements**

The aerogels were placed on the hot stage (Instec) as before (*cf.*, heat flux measurements) but with an infrared transparent KBr window instead of the copper block. Since the KBr lid cannot hold vacuum, the chamber was purged with pure nitrogen. An external port on the Fisher Scientific Nicolet iS50 FTIR was used to capture the emission. All experimental data was normalized to the response curve of a blackbody cavity (Infrared Systems Development Corporation IR-564).

### SEM/EDS measurements

SEM/EDS line-scans were collected using a TESCAN MIRA3 FEG SEM. Cross-section were made in aerogel monolith discs by creating a fracture surface near the middle of the disc after ALD.

### Uncertainty analysis

Uncertainty in the reported experimental quantities was evaluated based on propagation of the following errors: variance (using a t-distribution with a 95% confidence interval), instrument error and resolution error. All errors were assumed to be uncorrelated.

## Results & discussion

### Refractory aerogel composition

We fabricate transparent refractory aerogels by modifying the surface of silica aerogels with a single cycle of ALD, using trimethylaluminum (TMA) and deionized (DI) water as precursors (Figure 1 and Supplementary Note #1). Although ALD is generally considered to be a powerful method to coat porous materials,[26] it is challenging to uniformly coat aerogel monoliths because of their high surface area and tortuous, long, narrow pores. Furthermore, it is critical to deposit a conformal coating because superficial coatings onto the outer edges of the aerogel result in significant decreases in transmittance (Supplementary Note #2). To address this challenge, we recently develop a multidose-quasi-static-mode ALD process that enables conformal modification of ultra-high-aspect ratio pores (>60,000:1) with excellent precursor utilization.[21] Using this technique (see Methods), we conformally deposit a single cycle (~1-2 Å) of ALD on monolithic silica aerogel discs that are ~27 mm in diameter and ~3 mm in thickness (Supplementary Note #1).

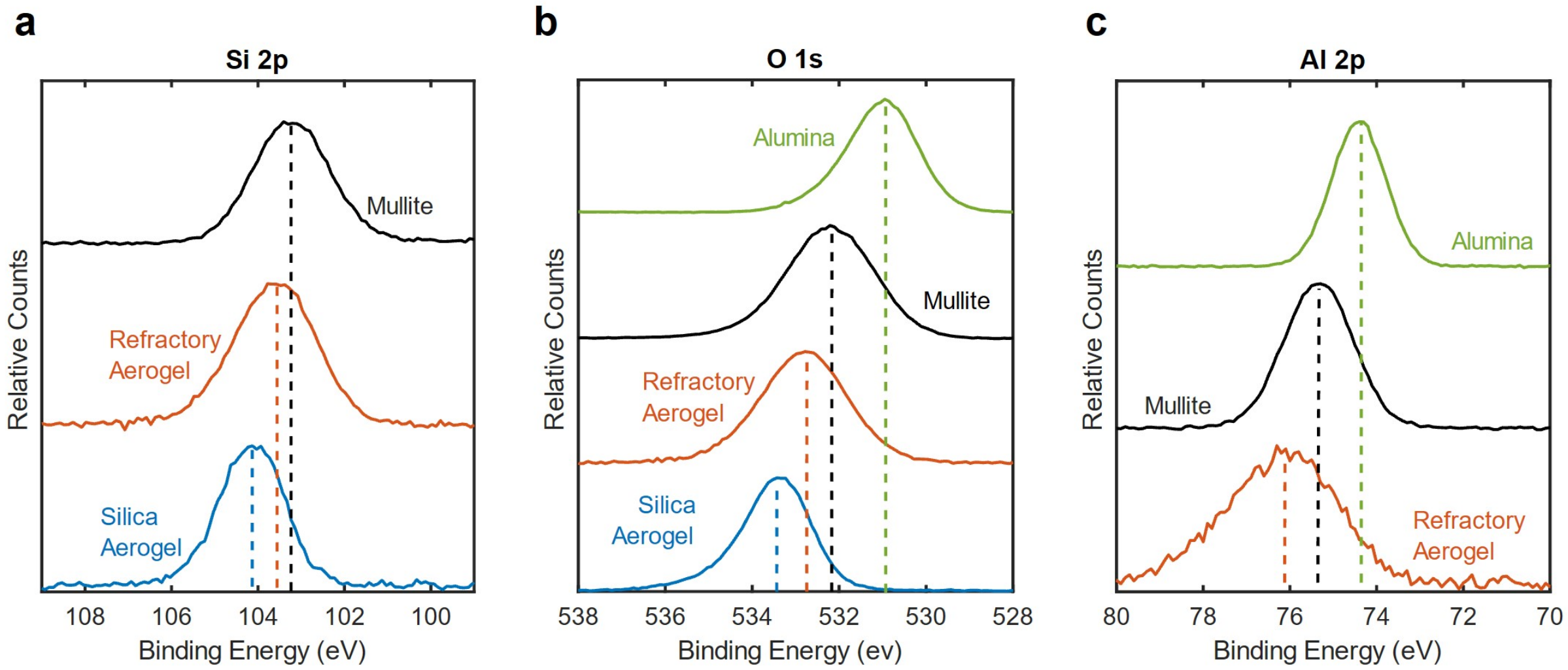

**Figure 2. Formation of an aluminum silicate aerogel chemistry.** XPS characterization of **(a)** Si 2p, **(b)** O 1s, and **(c)** Al 2p orbitals of the aerogels and standard references showing the formation of an aluminum silicate chemistry within the ALD-modified (refractory) aerogel.

The ALD process substantially alters the physical and chemical properties of the aerogels, largely due to their high surface area and small characteristic particle size. Following the ALD process, the aerogel density increases from ~175 kg/m$^3$ to ~230 kg/m$^3$, which represents a ~25% mass gain. Furthermore, energy dispersive x-ray spectroscopy (EDS) indicates that the composition changes from pure silicon oxide to a Si/Al ratio of ~3:1. We also note that EDS measurements indicate that the Al content is uniform throughout the aerogel volume, with no gradients in composition (and thus thickness) observed (Supplementary Note #1). X-ray diffraction (XRD) measurements reveal that the refractory aerogels are amorphous even after high-temperature annealing (Supplementary Note #3).

X-ray photoelectron spectroscopy (XPS) analysis was performed after the ALD process to investigate changes in binding environment. Although XPS is typically used to measure surface chemistry, the characteristic particle size (~6 nm) of the aerogels in this work is approximately twice the escape depth of photoelectrons in this material system (~2-3.5 nm). Therefore, XPS measurements can probe the overall aerogel chemistry, including bonding environments within the particle interior (see Methods for details). Figure 2 shows the measured Si 2p, O 1s, and Al 2p binding energies. XPS was also performed on reference mullite ($3Al_2O_3 \cdot 2SiO_2$) powder and bulk ALD alumina ($Al_2O_3$) samples. Figure 2a shows that the Si 2p peak of the refractory aerogel is in between that of mullite and the original silica aerogel. The shift of the Si 2p peak to a lower binding energy, as a result of the ALD modification, is consistent with previously reported shifts for increasing Al content in aluminum silicate compounds.[27] Similarly, the O 1s peak of refractory aerogel shifts to a lower binding energy (Figure 2b). Notably, the refractory aerogel O 1s peak is differentiated from the alumina peak and cannot be explained by a deconvolution of the spectrum into silica and alumina components. This differentiation relative to bulk alumina is also confirmed by the Al 2p peaks shown in Figure 2c.

These XPS data indicate that the ALD process results in the formation of a significant number of Al-O-Si moieties as opposed to the Al-O-Al moieties found in bulk alumina. This highlights a unique aspect of single-cycle ALD modification of an ultra-high surface area aerogel: these Al-O-Si linkages, which are the expected product species for a self-terminating reaction of TMA with hydroxylated amorphous silica,[28] significantly modify the overall composition of the aerogel because of their very high specific area (Table 1). Further studies are needed to determine the specific chemical coordination present; nonetheless, these data suggest that the binding environment at the outer surface of the refractory aerogel particles is analogous to that of known aluminum silicates.

## Improved thermal stability

Table 1 summarizes the improved thermal stability of the refractory aerogels by highlighting the physical properties of the aerogels before and after aging at 800°C for 14 days (see Methods and Supplementary Note #4). Furthermore, EDS analysis confirms that Al remains evenly distributed on a macroscale (monolith-level) throughout the aerogel following aging (Supplementary Note #5). Notably, after aging, the refractory aerogel has a larger specific surface area compared to its silica counterpart despite an initial decrease following ALD modification. Similarly, the aged refractory aerogel is less dense and has a higher porosity than its silica counterpart. These data indicate that the ALD modification stabilizes the highly mesoporous structure of the refractory aerogel. This improved structural stability results in a lower room-temperature thermal conductivity (by ~30%) in the aged refractory aerogel compared to its silica counterpart.

**Table 1.** Physical properties of aerogels before and after aggressive aging.
Detailed calculations of these properties are shown in Supplementary Note #5.

| Sample | Density ($kg/m^3$) | Porosity (%) | Surface area ($m^2/g$) | Thermal conductivity (mW/m/K) |
|---|---|---|---|---|
| Refractory, Initial | 230.5 ± 3.2 | 90.0 ± 1.2 | 662 | 18.6 ± 0.9 |
| Refractory, Aged | 333.0 ± 9.0 | 85.7 ± 2.3 | 503 | 30.6 ± 1.5 |
| Silica, Initial | 174.6 ± 2.4 | 92.1 ± 1.3 | 716 | 15.3 ± 0.8 |
| Silica, Aged | 387.6 ± 4.5 | 82.4 ± 1.0 | 481 | 41.4 ± 2.1 |

In addition to the discrete data points before and after 14 days of aggressive thermal aging, we report the linear shrinkage of the aerogels ($\Delta d/d$ where $d$ is the diameter of the aerogel disc) as a function of aging time at two different temperatures (Figure 3a). At 700ºC, both aerogels experience fast shrinkage within the first ~2 days, followed by slow shrinkage during days 2–10. After ~10 days at 700ºC, the linear shrinkage is only 8.6% for the refractory aerogel, compared to 16.9% for the silica aerogel. At 800ºC, the linear shrinkage for the refractory aerogel increases to 9.0% after one day, but it eventually stabilizes to a rate of 0.03%/day between days 17-20. In contrast, the silica aerogel continues to sinter at 800ºC with a rate of 0.29%/day (days 17-20). A comparison of the rates obtained during this period indicates that refractory aerogel has superior long-term stability and is likely suitable for operation at temperatures up to 800ºC.

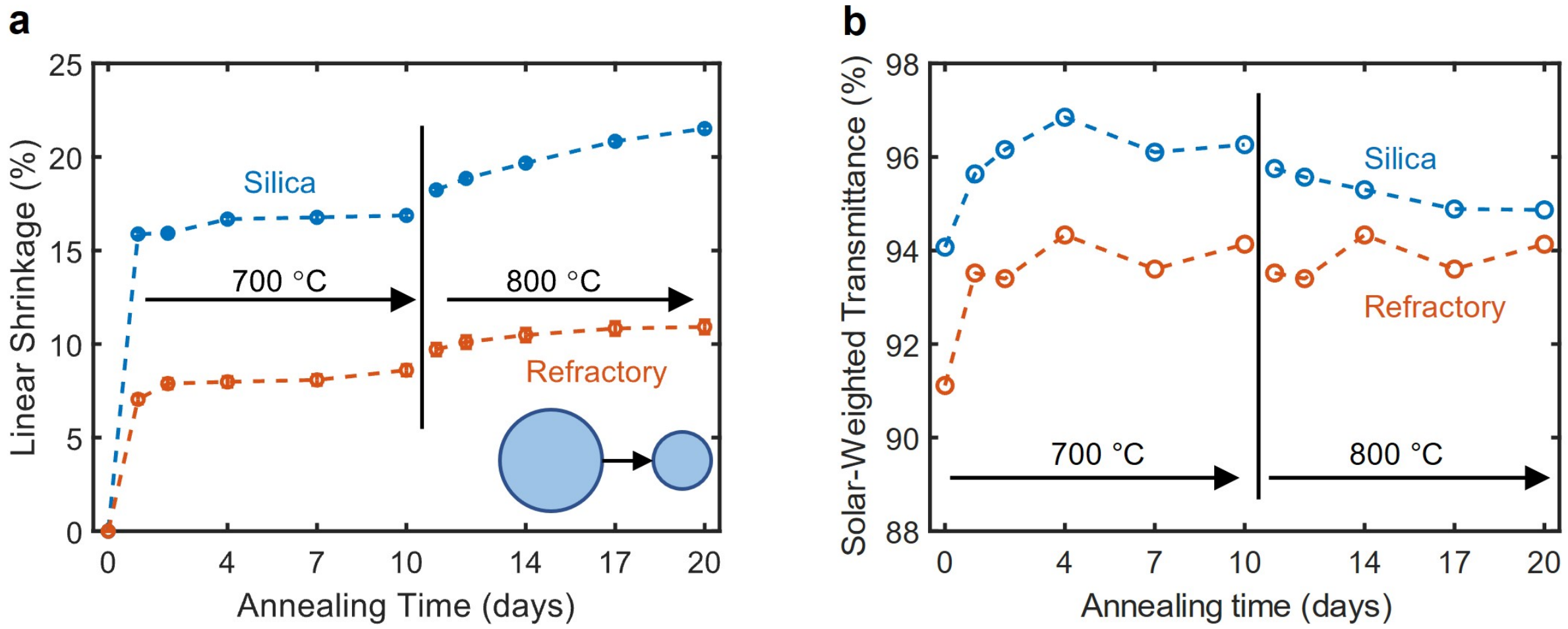

**Figure 3. Thermal stability of aerogels. (a)** Linear shrinkage as function of aging time demonstrating that the refractory aerogel is more stable at 800ºC, while the silica aerogel continues to densify. **(b)** Solar-weighted transmittance as a function of aging time, showing that that while the refractory aerogel has a lower initial transmittance by ~3%, the difference decreases to <1% by the end of the aggressive aging process.

The improved thermal stability resulting from the ALD modification can be attributed to several possible factors including kinetic, thermodynamic, and mechanical effects. Although other studies on alumina-silica aerogels have reported less shrinkage during drying and improved sintering resistance compared to silica aerogels,[20,29] the mechanisms are still not fully understood and may be different from the present work due to differences at a molecular scale. Specifically, Aravind *et al.* synthesized a homogenous alumina-silica aerogel *via* a sol-gel synthesis,[29] whereas the aerogels in this work were surface-modified. Sintering in silica aerogels is commonly attributed to a decrease in viscosity of the solid network at high temperatures which allows for structural relaxation and surface diffusion.[30] Within that theoretical framework, the formation of the aluminum silicate near the surface likely suppresses movement of the underlying silica. Analogous behavior has been observed in heterogeneous catalysis studies where thermally stable ALD "overcoats" slow down diffusion pathways that result in sintering of catalytic nanoparticles.[31] Nonetheless, further investigation is needed to identify the dominant mechanism leading to the observed high-temperature stability behavior in refractory aerogels.

The improvement in thermal stability is also reflected in solar-weighted transmittance measurements (Figure 3b). We note that the variation in the transmission data for the refractory aerogel is due to differences in relative humidity during measurements (Supplementary Note #6). After annealing at 700°C for 4 days, the transmittance of both aerogel samples increases slightly. Following this initial period, the refractory aerogel maintains a relatively constant solar-weighted transmittance. In contrast, the transmittance of the silica aerogel decreases as a result of sintering and the formation of larger particles and pores, which leads to increased light scattering.[17] While the difference in solar weighted transmittance is initially ~3%, the gap decreases to <1% by the end of the aging process. Overall, this time-resolved aging study confirms that the refractory aerogel maintains the high optical transmittance necessary for long-term use in CST applications.

**Optical and heat-insulating properties of aged refractory aerogels**

With the thermal stability of the refractory aerogels established above, we focus on the optical and thermal properties after aggressive aging and during operation at 700°C, as these characteristics dictate the performance of the aerogels in CST applications.

The spectral transmittance of the aged aerogels is shown in Figure 4a. The aged silica and refractory aerogels have nearly identical solar-weighted transmittances of 95% and 94%, respectively. The small decrease in transmittance in the refractory aerogel is mainly due to increased water absorption in the infrared and increased scattering at wavelengths <1 μm associated with a higher refractive index of the solid (Supplementary Note #6). As noted briefly before, this transmission penalty is small when compared to a superficial alumina coating applied onto the external surface of the aerogel monolith which decreases transmittance by >10% due to Fresnel reflections (Supplementary Note #2). This comparison highlights the importance of conformality in the ALD process.

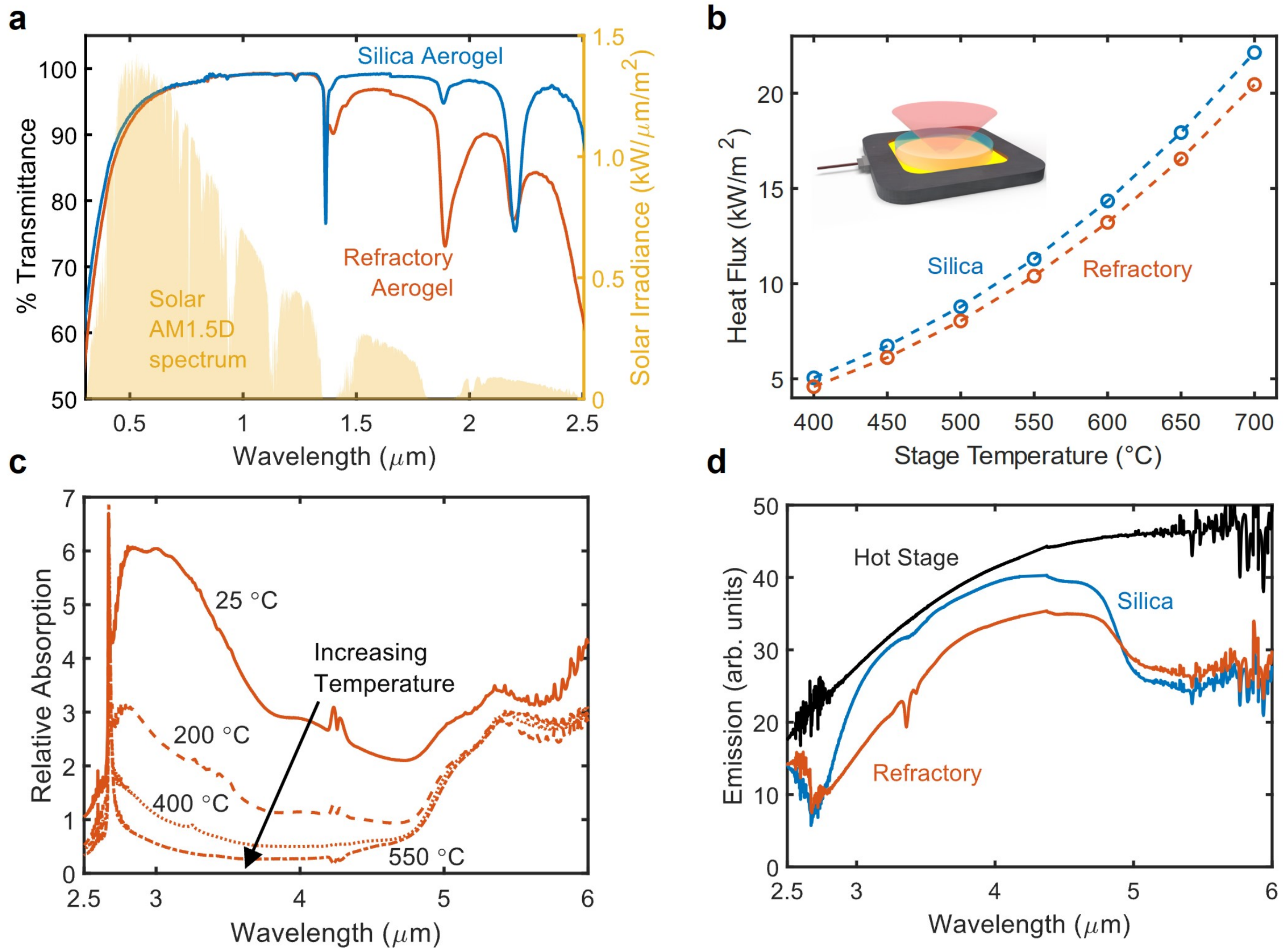

**Figure 4. Spectral optical and heat-transfer properties of aged refractory aerogels. (a)** Hemispherical transmittance data of aged aerogels. **(b)** Measured heat losses from the aerogel-covered SiC heater (*i.e.*, absorber) as a function of stage temperature. Error bars are omitted for clarity as they are all within 3% of the measured values. Inset: schematic showing emission from the hot stage before reaching the heat flux sensor or FTIR detector. The heat flux sensor or FTIR detector are not shown, but are above the silica aerogel. **(c)** Temperature-dependent DRIFTS showing the effects of adsorbed water on thermal opacity of refractory aerogels as a function of temperature. **(d)** Relative emission measurements for a 650°C absorber surface temperature in a nitrogen atmosphere. The refractory aerogel suppresses thermal radiation within the IR transparency window of the silica aerogel (3-5 μm).

We assess the thermal insulating performance of the aged refractory aerogel by measuring the overall heat loss rate ($Q_{loss}$) as a function of absorber temperature in an evacuated environment, following a previously reported procedure.[16] The aerogel is placed on a relatively high emissivity silicon carbide surface, designated as the solar absorber. The heat flux reaching a black sensor in proximity of the cold-side of the aerogel is measured as the temperature of the absorber is gradually increased. We note that the heat flux increases only marginally after the hot stage is painted with a commercial high emissivity paint (Supplementary Note #7).

Figure 4b shows that the aged refractory aerogel suppresses heat losses more effectively than its silica counterpart. The difference in heat loss at these high temperatures can be explained by two effects: (1) the refractory aerogel conducts less heat through its solid network, which is consistent with the room-temperature thermal conductivity measurements in Table 1, and (2) there is a reduction in radiative heat transfer associated with an increase in IR absorption within the refractory aerogels. The ALD modification introduces new absorption modes at longer wavelengths due to the stretching (~11.5 μm) and vibrational (~16 μm) modes typical of alumina-silica chemistries (Supplementary Note #8). However, absorption at these long wavelengths is not expected to have a significant effect on radiative heat transfer within the aerogel because only a small fraction of the emissive power of a blackbody lies above 10 μm at relevant temperatures (e.g., ~10% at 700ºC). Instead, the increased thermal opacity of the aged refractory aerogel is most likely due to adsorbed water, as well as surface hydroxyl groups, which absorb thermal emission within the mid-IR transparency window of pure silica aerogels (3-5 μm). To study this effect, we performed temperature-dependent diffuse reflectance FTIR (DRIFTS) measurements (Figure 4c) and thermogravimetric analysis (TGA) on powdered samples (see Methods and Supplementary Note #8). DRIFTS and TGA data indicate the presence of water up to 400°C in the aged refractory aerogel. In contrast, water is not present in the aged silica aerogel due to surface condensation of silanol groups that occurs during annealing.[32] Notably, the refractory aerogel retains its hydrophilic nature even after prolonged aging above 700°C. This behavior is most likely due to the strong

Lewis acidity of unsaturated (triply coordinated) surface aluminum sites which strongly adsorb water (-131 kJ/mol).[33] Furthermore, computational studies show that terminating these sites with hydroxyl groups produces a more stable tetrahedral coordination.[28,34] Although high temperatures may cause eventual dehydroxylation, exposure to humid air can regenerate hydrophilic groups – a phenomenon that we observe in the refractory aerogel (Supplementary Note #8).[35] This temperature-dependent water adsorption causes the refractory aerogel to absorb more IR radiation at colder temperatures (Figure 4c). These results are consistent with ALD literature showing that even a single cycle of ALD can significantly change the hydrophobicity/philicity of porous substrates.[36] The effective thermal emittance, $\varepsilon_{eff}$ (defined as the ratio of heat losses to the blackbody emissive power at the absorber temperature), decreases from 0.44 to 0.40 at 700°C due to the ALD modification.

We further confirm the increase in mid-IR absorption via FTIR thermal emission measurements (Figure 4d). In this experiment, aerogels were placed on the hot stage as before but with an infrared transparent KBr window instead of the opaque heat flux sensor. Emission through the silica aerogel in the 3–5 µm transparency window is nearly equal to the emission from the hot stage, confirming that the silica aerogel is almost fully transparent over that wavelength range. In contrast, the refractory aerogel suppresses emission within the 3–5 µm wavelength range. The presence of a temperature gradient along the aerogel thickness likely allows some water to be retained in the colder regions of the refractory aerogel, which acts to block outgoing emission. Modeling results suggest that the temperature of cold side of the silica aerogel is as low as 275°C when placed on a 700°C absorber (Supplementary Note #9). At this temperature, the refractory aerogel retains enough adsorbed water (Figure 4c, Supplementary Note #8) to suppress radiative losses.

Overall, we find that the ALD modification has a minimal negative impact on solar transmittance because of its conformality, while having a positive impact on thermal opacity. In particular, the Al-rich surface of the refractory aerogels has surface sites that strongly adsorb water and appear to be robust at high temperatures. This mechanism represents a new approach for selectively enhancing thermal opacity at high temperatures, which appears to be complementary to existing strategies, including our prior work on plasmon-enhanced greenhouse selectivity (PEGS) using embedded ITO nanoparticles.[16]

**Receiver performance**

The performance of refractory aerogels is compared to state-of-the-art selective surfaces based on receiver efficiency ($\eta$), as described by Equation 1.

$$\eta = \alpha\tau - \frac{Q_{loss}}{CG_s} \tag{1}$$

Here, $\alpha$ is the measured solar-weighted absorptance of the absorber, $\tau$ is the solar-weighted transmittance of the aerogel ($\tau$ is set to 1 for selective surfaces), $Q_{loss}$ is the measured rate of temperature-dependent heat loss (*cf.*, Figure 4b for refractory aerogels), $C$ is the solar concentration ratio, and $G_s$ is the standard AM1.5D incident solar power (*i.e.*, 1,000 W/m$^2$). There are several approaches for achieving a near-blackbody absorber, such as using multi-walled carbon nanotubes and macroscale cavities with large aspect ratios.[37–39] Since aerogels can integrate with such absorbers, we thus set $\alpha$ equal to 1 in the case of aerogels.

The improvement demonstrated by the refractory aerogel has the potential to increase the solar-to-electricity conversion efficiency by ~10% (absolute) by enabling higher efficiency $sCO_2$ power cycles while maintaining a high receiver efficiency (Supplementary Note #10). Furthermore, aerogels exhibit pressure-independent thermal resistance below 10 mbar,[40] which can enable lower-cost receiver designs that do not require high vacuum levels and/or help maintain performance over time by avoiding issues related to vacuum breakdown. In this analysis, we exclude all selective surfaces whose emittance was measured below 700°C. Although selective surfaces are commonly annealed at elevated temperatures to test their stability, the emissivity measurements are often performed at or near room temperature. We exclude these reports from this analysis because low temperature measurements significantly overestimate the high temperature performance of selective absorbers.[41,42] However, a summary of the spectral selectivity ($\alpha_{eff}/\varepsilon_{eff}$) of selective surfaces tested above 500°C is tabulated in Supplementary Note #11.

Figure 5 shows the receiver efficiency of selective surfaces and TIMs using concentration ratios of 50 and 100 suns. The refractory aerogel outperforms all reports of systems measured at ≥700°C to date.[16,42,43] The refractory aerogel is more transparent (94%) than selective surfaces are absorptive, leading to a higher receiver efficiency (Equation 1).

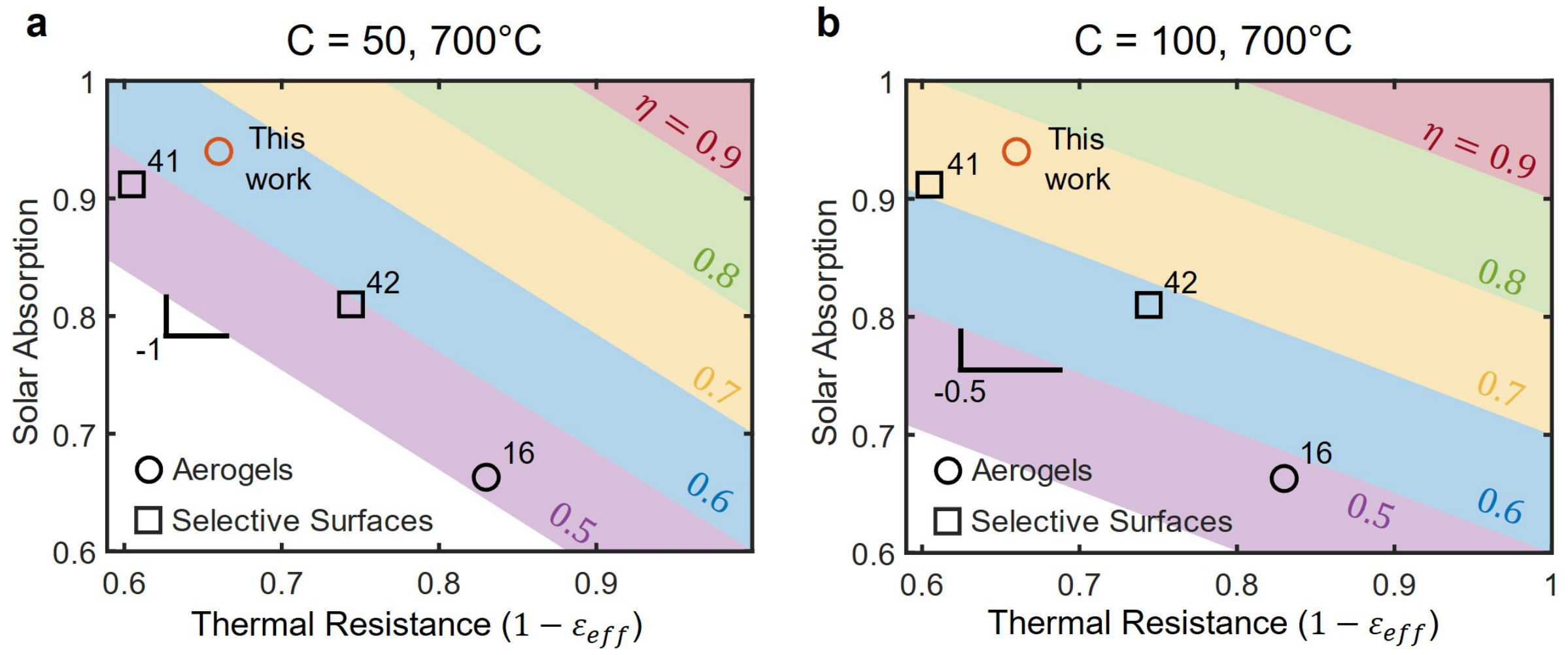

**Figure 5. Calculated receiver efficiency at 700°C.** The refractory aerogel achieves a higher receiver efficiency at concentration ratios ($C$) of **(a)** 50 and **(b)** 100 suns than all materials and surfaces tested at ≥700ºC to date.

Figure 5 also identifies a key takeaway for the future development of solar-selective materials and receivers. At 700°C, the emissive power of a blackbody is ~50 kW/m$^2$. Therefore, at a concentration ratio of ~50, the incoming solar flux equals the outgoing blackbody emissive power. Increasing solar transmittance by 1% or decreasing thermal emittance by 1% will improve efficiency by 1%. This explains why the slope of the efficiency lines in Figure 5a is –1. At concentration ratios below 50 suns, it is more important to reduce the thermal emittance, while the opposite is true for concentrations above 50 suns. The acceptable tradeoff is given by $\Delta\varepsilon_{eff}/\Delta\tau \geq C/50$, where $\Delta$ represents a decrease in the corresponding property. Future work on aerogels for solar trough collectors (C~80) should thus focus on methods to reduce the thermal emittance by at least 1.6 units for each unit lost in solar transmission. Meanwhile, aerogels in lower irradiance systems can prioritize more aggressive thermal insulating strategies over solar transmittance, including PEGS[16] or doping of the silica network.[44] Furthermore, the thickness of the aerogel can also be optimized to fit the needs of the solar thermal plant. Unlike selective surfaces which may require fundamental changes to structure or patterning,[43] the refractory aerogel can simply be made thicker or thinner depending on the application. Alternatively, the aerogels could be stacked on top of each other to provide more insulation if necessary.[12] Beyond concentrated solar power, refractory aerogels have the potential to impact a broad range of technologies where spectral control of solar and thermal radiation is important, including in hybrid photovoltaic/thermal systems,[45–48] infrared light sources,[49,50] radiative heat engines,[1,51–53] and solar thermochemical processes.[54,55]

## Conclusions

This work presents an approach to overcome temperature-related degradation of aerogels while preserving their exceptional optical and heat-insulating properties, thus overcoming a major barrier to deploying aerogels in next-generation CST plants. Conformal modification of silica aerogel monoliths via ALD significantly increases the overall percentage of aluminum (Si/Al ratio ~ 3:1) and produces an aluminum silicate phase that is likely situated near the surface. The resulting refractory aerogel is only slightly less transparent than silica aerogels (94% to 95% after aging), while demonstrating improved high-temperature stability and thermal opacity. At 800°C, the linear shrinkage rate is 0.03% per day for the refractory aerogel compared to 0.29% for the silica aerogel control, demonstrating improved structural stability. Furthermore, among selective surfaces and TIMs that have been tested at ≥700°C, the refractory aerogel can achieve the highest receiver efficiency to date. Given the stable performance at high temperatures, transparent refractory aerogel insulation has the potential to unlock the development of modular, high-efficiency solar thermal plants which could provide on-demand renewable energy at lower costs than existing approaches.[56,57]

## Competing financial interests

The authors have filed a provisional patent application related to this work.

## Acknowledgements

This work was supported by U.S. Department of Energy's Office of Energy Efficiency and Renewable Energy (EERE) under the Solar Energy Technologies Office (SETO) Award Number DE-EE0008526 and DE-EE0009376. The view expressed herein do not necessarily represent the view of the U.S. Department of Energy or the United States Government. This work is supported by the National Science Foundation under Grant No. 1751590. Z.J.B. and A.J.G. acknowledge support from the National Science Foundation Graduate Research Fellowship Program under Grant No. DGE-1256260. Z.J.B. also acknowledges support from the University of Michigan Dow Sustainability Doctoral Research Fellowship and the Mistletoe Research Fellowship. The SEM/EDS work was performed at the Michigan Center for Materials Characterization and aerogel critical point drying was performed at the Lurie Nanofabrication Facility, which are supported by the College of Engineering at University of Michigan. We acknowledge Alexander J. Hill for contributions on nitrogen sorption measurements.

## Author Information

Corresponding Authors
*E-mail: ndasgupt@umich.edu
*E-mail: alenert@umich.edu

ORCID
Neil P. Dasgupta: 0000-0002-5180-4063
Andrej Lenert: 0000-0002-1142-6627

## Author Contributions

Z.J.B. and A.J.G. contributed equally to this work.

## References

(1) Seyf, H. R.; Henry, A. Thermophotovoltaics: A Potential Pathway to High Efficiency Concentrated Solar Power. *Energy Environ. Sci.* **2016**, *9* (8), 2654–2665.
(2) Mehos, M.; Turchi, C.; Vidal, J.; Wagner, M.; Ma, Z.; Ho, C.; Kolb, W.; Andraka, C.; Kruizenga, A. Concentrating Solar Power Gen3 Demonstration Roadmap. *Nrel/Tp-5500-67464* **2017**, No. January, 1–140.
(3) Ru Ng, A. Y.; Boruah, B.; Chin, K. F.; Modak, J. M.; Soo, H. Sen. Photoelectrochemical Cells for Artificial Photosynthesis: Alternatives to Water Oxidation. *ChemNanoMat* **2020**, *6* (2), 185–203.
(4) Turchi, C. S.; Ma, Z.; Neises, T. W.; Wagner, M. J. Thermodynamic Study of Advanced Supercritical Carbon Dioxide Power Cycles for Concentrating Solar Power Systems. *J. Sol. Energy Eng.* **2013**, *135* (4), 1–7.
(5) Lavine, A. S.; Lovegrove, K. M.; Jordan, J.; Anleu, G. B.; Chen, C.; Aryafar, H.; Lavine, A. S.; Lovegrove, K. M.; Jordan, J.; Bran, G.; Chen, C.; Aryafar, H.; Sepulveda, A. Thermochemical Energy Storage with Ammonia: Aiming for the SunShot Cost Target. In *AIP Conference Proceedings*; 2016; Vol. 1734, p 050028.
(6) US Department of Energy. Solar Research Spotlight: Concentrating Solar Thermal Power. *Off. Energy Effic. Renew. Energy* **2018**.
(7) Thiel, G. P.; Stark, A. K. To Decarbonize Industry, We Must Decarbonize Heat. *Joule* **2021**, No. 2020, 1–20.
(8) Tagle-Salazar, P. D.; Nigam, K. D. P.; Rivera-Solorio, C. I. Parabolic Trough Solar Collectors: A General Overview of Technology, Industrial Applications, Energy Market, Modeling, and Standards. *Green Processing and Synthesis*. De Gruyter Open Ltd January 2020, pp 595–649.
(9) Kincaid, N.; Mungas, G.; Kramer, N.; Wagner, M.; Zhu, G. An Optical Performance Comparison of Three Concentrating Solar Power Collector Designs in Linear Fresnel, Parabolic Trough, and Central Receiver. *Appl. Energy* **2018**, *231* (September), 1109–1121.
(10) Zhang, K.; Hao, L.; Du, M.; Mi, J.; Wang, J.; Meng, J. A Review on Thermal Stability and High Temperature Induced Ageing Mechanisms of Solar Absorber Coatings. *Renew. Sustain. Energy Rev.* **2017**, *67*, 1282–1299.
(11) Xu, K.; Du, M.; Hao, L.; Mi, J.; Yu, Q.; Li, S. A Review of High-Temperature Selective Absorbing Coatings for Solar Thermal Applications. *J. Mater.* **2020**, *6* (1), 167–182.
(12) Zhao, L.; Bhatia, B.; Yang, S.; Strobach, E.; Weinstein, L. A.; Cooper, T. A.; Chen, G.; Wang, E. N. Harnessing Heat beyond 200 °c from Unconcentrated Sunlight with Nonevacuated Transparent Aerogels. *ACS Nano* **2019**, *13* (7), 7508–7516.
(13) Zhao, L.; Bhatia, B.; Zhang, L.; Strobach, E.; Leroy, A.; Yadav, M. K.; Yang, S.; Cooper, T. A.; Weinstein, L. A.; Modi, A.; Kedare, S. B.; Chen, G.; Wang, E. N. A Passive High-Temperature High-Pressure Solar Steam Generator for Medical Sterilization. *Joule* **2020**, *4* (12), 2733–2745.
(14) Caps, R.; Fricke, J. Infrared Radiative Heat Transfer in Highly Transparent Silica Aerogel. *Sol. Energy* **1986**, *36* (4), 361–364.
(15) Zeng, S. Q.; Hunt, A.; Greif, R. Theoretical Modeling of Carbon Content to Minimize Heat

Transfer in Silica Aerogel. **1995**, *186*, 271–277.
(16) Berquist, Z. J.; Turaczy, K. K.; Lenert, A. Plasmon-Enhanced Greenhouse Selectivity for High-Temperature Solar Thermal Energy Conversion. *ACS Nano* **2020**, *14* (10), 12605–12613.
(17) Strobach, E.; Bhatia, B.; Yang, S.; Zhao, L.; Wang, E. N. High Temperature Stability of Transparent Silica Aerogels for Solar Thermal Applications. *APL Mater.* **2019**, *7* (8).
(18) Venkateswara Rao, A.; Bhagat, S. D.; Hirashima, H.; Pajonk, G. M. Synthesis of Flexible Silica Aerogels Using Methyltrimethoxysilane (MTMS) Precursor. *J. Colloid Interface Sci.* **2006**, *300* (1), 279–285.
(19) Maleki, H.; Durães, L.; Portugal, A. An Overview on Silica Aerogels Synthesis and Different Mechanical Reinforcing Strategies. *J. Non. Cryst. Solids* **2014**, *385*, 55–74.
(20) Himmel, B.; Gerber, T.; Bürger, H.; Holzhüter, G.; Olbertz, A. Structural Characterization of SiO2-Al2O3 Aerogels. *J. Non. Cryst. Solids* **1995**, *186*, 149–158.
(21) Gayle, A. J.; Berquist, Z. J.; Chen, Y.; Hill, A. J.; Hoffman, J. Y.; Ashley, R.; Lenert, A.; Dasgupta, N. P. Tunable Atomic Layer Deposition into Ultra-High-Aspect-Ratio ( > 60 , 000 : 1 ) Aerogel Monoliths Enabled by Transport Modeling. *Chem. Mater.* **2021**, No. 33.
(22) ASTM Standard C71-12, 2021, " Standard Terminology Relating to Refractories, " ASTM International, West Conshohocken, PA, 2021. **2021**, 2021.
(23) Guler, U.; Boltasseva, A.; Shalaev, V. M. Refractory Plasmonics. *Science (80-. ).* **2014**, *344* (6181), 263–264.
(24) Dasgupta, N. P.; Mack, J. F.; Langston, M. C.; Bousetta, A.; Prinz, F. B. Design of an Atomic Layer Deposition Reactor for Hydrogen Sulfide Compatibility. *Rev. Sci. Instrum.* **2010**, *81* (4), 044102.
(25) Zheng, Q.; Kaur, S.; Dames, C.; Prasher, R. S. Analysis and Improvement of the Hot Disk Transient Plane Source Method for Low Thermal Conductivity Materials. *Int. J. Heat Mass Transf.* **2020**, *151*.
(26) Elam, J. W.; Xiong, G.; Han, C. Y.; Wang, H. H.; Birrell, J. P.; Welp, U.; Hryn, J. N.; Pellin, M. J.; Baumann, T. F.; Poco, J. F.; Jr, J. H. S. Atomic Layer Deposition for the Conformal Coating of Nanoporous Materials. *J. Nanomater.* **2006**, *2006*, 1–5.
(27) Stoch, J.; Lercher, J.; Ceckiewicz, S. Correlations between XPS Binding Energies and Composition of Aluminasilicate and Phosphate Molecular Sieves. *Zeolites* **1992**, *12* (1), 81–85.
(28) Sandupatla, A. S.; Alexopoulos, K.; Reyniers, M. F.; Marin, G. B. DFT Investigation into Alumina ALD Growth Inhibition on Hydroxylated Amorphous Silica Surface. *J. Phys. Chem. C* **2015**, *119* (32), 18380–18388.
(29) Aravind, P. R.; Mukundan, P.; Krishna Pillai, P.; Warrier, K. G. K. Mesoporous Silica-Alumina Aerogels with High Thermal Pore Stability through Hybrid Sol-Gel Route Followed by Subcritical Drying. *Microporous Mesoporous Mater.* **2006**, *96* (1–3), 14–20.
(30) Brinker, C. J.; Scherer, G. W. *Sol-Gel Science*; Academic Press, Inc.: San Diego, CA, 1990.
(31) Oneill, B. J.; Jackson, D. H. K.; Lee, J.; Canlas, C.; Stair, P. C.; Marshall, C. L.; Elam, J. W.; Kuech, T. F.; Dumesic, J. A.; Huber, G. W. Catalyst Design with Atomic Layer Deposition. *ACS Catal.* **2015**, *5* (3), 1804–1825.
(32) Malfait, W. J.; Zhao, S.; Verel, R.; Iswar, S.; Rentsch, D.; Fener, R.; Zhang, Y.; Milow, B.; Koebel, M. M. Surface Chemistry of Hydrophobic Silica Aerogels. *Chem. Mater.* **2015**, *27* (19), 6737–6745.
(33) Digne, M.; Sautet, P.; Raybaud, P.; Euzen, P.; Toulhoat, H. Hydroxyl Groups on γ-Alumina Surfaces: A DFT Study. *J. Catal.* **2002**, *211* (1), 1–5.
(34) Digne, M.; Sautet, P.; Raybaud, P.; Euzen, P.; Toulhoat, H. Use of DFT to Achieve a Rational Understanding of Acid-Basic Properties of γ-Alumina Surfaces. *J. Catal.* **2004**, *226* (1), 54–68.
(35) Petrik, N. G.; Huestis, P. L.; Laverne, J. A.; Aleksandrov, A. B.; Orlando, T. M.; Kimmel, G. A. Molecular Water Adsorption and Reactions on α-Al2O3(0001) and α-Alumina Particles. *J. Phys. Chem. C* **2018**, *122* (17), 9540–9551.
(36) Li, Y.; Chen, L.; Wooding, J. P.; Zhang, F.; Lively, R. P.; Ramprasad, R.; Losego, M. D.

Controlling Wettability, Wet Strength, and Fluid Transport Selectivity of Nanopaper with Atomic Layer Deposited (ALD) Sub-Nanometer Metal Oxide Coatings. *Nanoscale Adv.* **2020**, *2* (1), 356–367.
(37) Yang, Z. P.; Ci, L.; Bur, J. A.; Lin, S. Y.; Ajayan, P. M. Experimental Observation of an Extremely Dark Material Made by a Low-Density Nanotube Array. *Nano Lett.* **2008**, *8* (2), 446–451.
(38) Lenert, A.; Bierman, D. M.; Nam, Y.; Chan, W. R.; Celanović, I.; Soljačić, M.; Wang, E. N. A Nanophotonic Solar Thermophotovoltaic Device. *Nat. Nanotechnol.* **2015**, *10* (6), 563–563.
(39) Lin, H.; Sturmberg, B. C. P.; Lin, K. Te; Yang, Y.; Zheng, X.; Chong, T. K.; de Sterke, C. M.; Jia, B. A 90-Nm-Thick Graphene Metamaterial for Strong and Extremely Broadband Absorption of Unpolarized Light. *Nat. Photonics* **2019**, *13* (4), 270–276.
(40) He, Y. L.; Xie, T. Advances of Thermal Conductivity Models of Nanoscale Silica Aerogel Insulation Material. *Appl. Therm. Eng.* **2015**, *81*, 28–50.
(41) Echániz, T.; Setién-Fernández, I.; Pérez-Sáez, R. B.; Prieto, C.; Galindo, R. E.; Tello, M. J. Importance of the Spectral Emissivity Measurements at Working Temperature to Determine the Efficiency of a Solar Selective Coating. *Sol. Energy Mater. Sol. Cells* **2015**, *140*, 249–252.
(42) Wang, H.; Alshehri, H.; Su, H.; Wang, L. Design, Fabrication and Optical Characterizations of Large-Area Lithography-Free Ultrathin Multilayer Selective Solar Coatings with Excellent Thermal Stability in Air. *Sol. Energy Mater. Sol. Cells* **2018**, *174* (June 2017), 445–452.
(43) Rinnerbauer, V.; Lenert, A.; Bierman, D. M.; Yeng, Y. X.; Chan, W. R.; Geil, R. D.; Senkevich, J. J.; Joannopoulos, J. D.; Wang, E. N.; Soljačič, M.; Celanovic, I. Metallic Photonic Crystal Absorber-Emitter for Efficient Spectral Control in High-Temperature Solar Thermophotovoltaics. *Adv. Energy Mater.* **2014**, *4* (12), 1–10.
(44) Yoshimaru, M.; Koizumi, S.; Shimokawa, K. Structure of Fluorine-Doped Silicon Oxide Films Deposited by Plasma-Enhanced Chemical Vapor Deposition. *J. Vac. Sci. Technol. A Vacuum, Surfaces, Film.* **1997**, *15* (6), 2908–2914.
(45) Taylor, R. A.; Otanicar, T.; Rosengarten, G. Nanofluid-Based Optical Filter Optimization for PV/T Systems. *Light Sci. Appl.* **2012**, *1* (OCTOBER), 1–7.
(46) Bierman, D. M.; Lenert, A.; Wang, E. N. Spectral Splitting Optimization for High-Efficiency Solar Photovoltaic and Thermal Power Generation. *Appl. Phys. Lett.* **2016**, *109* (24).
(47) Weinstein, L. A.; McEnaney, K.; Strobach, E.; Yang, S.; Bhatia, B.; Zhao, L.; Huang, Y.; Loomis, J.; Cao, F.; Boriskina, S. V.; Ren, Z.; Wang, E. N.; Chen, G. A Hybrid Electric and Thermal Solar Receiver. *Joule* **2018**, *2* (5), 962–975.
(48) Yu, Z. J.; Fisher, K. C.; Meng, X.; Hyatt, J. J.; Angel, R. P.; Holman, Z. C. GaAs/Silicon PVMirror Tandem Photovoltaic Mini-Module with 29.6% Efficiency with Respect to the Outdoor Global Irradiance. *Prog. Photovoltaics Res. Appl.* **2019**, *27* (5), 469–475.
(49) Baranov, D. G.; Xiao, Y.; Nechepurenko, I. A.; Krasnok, A.; Alù, A.; Kats, M. A. Nanophotonic Engineering of Far-Field Thermal Emitters. *Nat. Mater.* **2019**, *18* (9), 920–930.
(50) Xu, J.; Raman, A. P. Broadband Directional Control of Thermal Emission. *Science (80-. ).* **2021**, *397* (April), 393–397.
(51) Chan, W. R.; Stelmakh, V.; Ghebrebrhan, M.; Soljačić, M.; Joannopoulos, J. D.; Čelanović, I. Enabling Efficient Heat-to-Electricity Generation at the Mesoscale. *Energy Environ. Sci.* **2017**, *10* (6), 1367–1371.
(52) Wang, Y.; Liu, H.; Zhu, J. Solar Thermophotovoltaics: Progress, Challenges, and Opportunities. *APL Mater.* **2019**, *7* (8).
(53) Horiuchi, N. Efficient Thermophotovoltaics. *Nat. Photonics* **2020**, *14* (2), 66–66.
(54) Marxer, D.; Furler, P.; Takacs, M.; Steinfeld, A. Solar Thermochemical Splitting of CO2 into Separate Streams of CO and O2 with High Selectivity, Stability, Conversion, and Efficiency. *Energy Environ. Sci.* **2017**, *10* (5), 1142–1149.
(55) Tembhurne, S.; Nandjou, F.; Haussener, S. A Thermally Synergistic Photo-Electrochemical Hydrogen Generator Operating under Concentrated Solar Irradiation. *Nat. Energy* **2019**, *4* (5),

399–407.
(56) Yadav, D.; Banerjee, R. A Review of Solar Thermochemical Processes. *Renew. Sustain. Energy Rev.* **2016**, *54*, 497–532.
(57) Carrillo, A. J.; González-Aguilar, J.; Romero, M.; Coronado, J. M. Solar Energy on Demand: A Review on High Temperature Thermochemical Heat Storage Systems and Materials. *Chem. Rev.* **2019**, *119* (7), 4777–4816.